\documentclass[a4paper,amsmath,amssymb]{jpconf}
\usepackage{graphicx}
\usepackage{iwsmihead}
\usepackage{amsmath, amsthm, amssymb}
\usepackage{hyperref}

\begin{document}

\title{Generalized minimum dominating set and application in automatic text summarization}

\author{Yi-Zhi Xu and Hai-Jun Zhou}

\address{State Key Laboratory of Theoretical Physics, Institute of Theoretical
  Physics, Chinese Academy of Sciences, Zhong-Guan-Cun East Road 55, 
  Beijing 100190, China
}

\ead{xyz@itp.ac.cn, zhouhj@itp.ac.cn}

\begin{abstract}
  For a graph formed by vertices and weighted edges, 
  a generalized minimum dominating set (MDS) is a vertex set of smallest 
  cardinality such that the summed weight of edges from each outside vertex
  to vertices in this set is equal to or larger than certain threshold value. 
  This generalized MDS problem reduces to the conventional MDS problem in
  the limiting case of all the edge weights being equal to the threshold value. 
  We treat the generalized MDS problem in the present paper by a
  replica-symmetric spin glass theory and derive a set of belief-propagation
  equations. As a practical application we consider the problem of extracting
  a set of sentences that best summarize a given input text document.
  We carry out a preliminary test of the statistical physics-inspired method
  to this automatic text summarization problem.
\end{abstract}

\section{Introduction}
\label{sec.intr}

Minimum dominating set (MDS) is a well-known concept
in the computer science community
(see review \cite{Haynes-Hedetniemi-Slater-1998}). For a given
graph, a MDS is just a minimum-sized vertex set such that either
a vertex belongs to this set or at least one of its neighbors
belongs to this set.  In the last few years 
researchers from the statistical physics community also got 
quite interested in this concept, as it is closely related to various
network problems such as
network monitoring, network control, infectious disease suppression,
and resource allocation (see, for example, 
\cite{Echenique-etal-2005,DallAsta-Pin-Ramezanpour-2009,DallAsta-Pin-Ramezanpour-2011,Yang-Wang-Motter-2012,Molnar-etal-2013,Nacher-Akutsu-2013,Takaguchi-Hasegawa-Yoshida-2014,Wuchty-2014,Wang-etal-2014}
and review \cite{Liu-Barabasi-2015}).
Constructing an exact MDS for a large graph is, generally speaking, an
extremely difficult task and it is very likely that no complete algorithm is
capable of solving it in an efficient way. On the other hand, by mapping
the MDS problem into a spin glass system with local many-body constraints and
then treating it by statistical-physics methods, one can estimate with high
empirical confidence the sizes of minimum dominating sets for single graph
instances \cite{Zhao-Habibulla-Zhou-2015,Habibulla-Zhao-Zhou-2015}. One can 
also construct close-to-minimum dominating sets quickly through a 
physics-inspired heuristic algorithm 
\cite{Zhao-Habibulla-Zhou-2015,Habibulla-Zhao-Zhou-2015},
which might be important for many practical applications.

In the present work we extend the statistical-physics approach of
\cite{Zhao-Habibulla-Zhou-2015,Habibulla-Zhao-Zhou-2015} to edge-weighted
graphs and study a generalized minimum dominating set problem. Our work
is motivated by a practical knowledge-mining problem: extracting a set of
sentences to best summarize one or more input text documents 
\cite{Mani-1999,Shen-Li-2010}. We consider a general graph of vertices
and edges, each edge connecting two different vertices and bearing one weight
or a pair of weights (see Fig.~\ref{fig:exmpGM}). In the context of text
summarization, a vertex represents a sentence of some text documents and an
edge weight is the similarity between two sentences. Various data-clustering
problems can also be represented as weighted graphs. Given such a weighted
graph, our task is then to construct a minimum-cardinality set $\Gamma_0$ of
vertices such that if a vertex $i$ is not included in $\Gamma_0$, the summed
weight of the edges from $i$ to vertices in $\Gamma_0$ must reach at least 
certain threshold value $\theta$. The set $\Gamma_0$ is referred to as
a (generalized) MDS.

We introduce a spin glass model for this generalized MDS problem in
Sec.~\ref{sec.SpGlas} and then describe a replica-symmetric (RS) mean field
theory in Sec.~\ref{sec.mft}. A message-passing algorithm BPD
(belief-propagation guided decimation) is outlined in Sec.~\ref{sec.BPD},
and is then applied to the automatic text summarization problem in
Sec.~\ref{sec.app}. We conclude this work in Sec.~\ref{sec.conc} and discuss
a way of modifying the spin glass model for better treating the text 
summarization problem.

\section{Constraints and a spin glass model}
\label{sec.SpGlas}

\begin{figure}
  \begin{center}
    \includegraphics[width=0.4\textwidth]{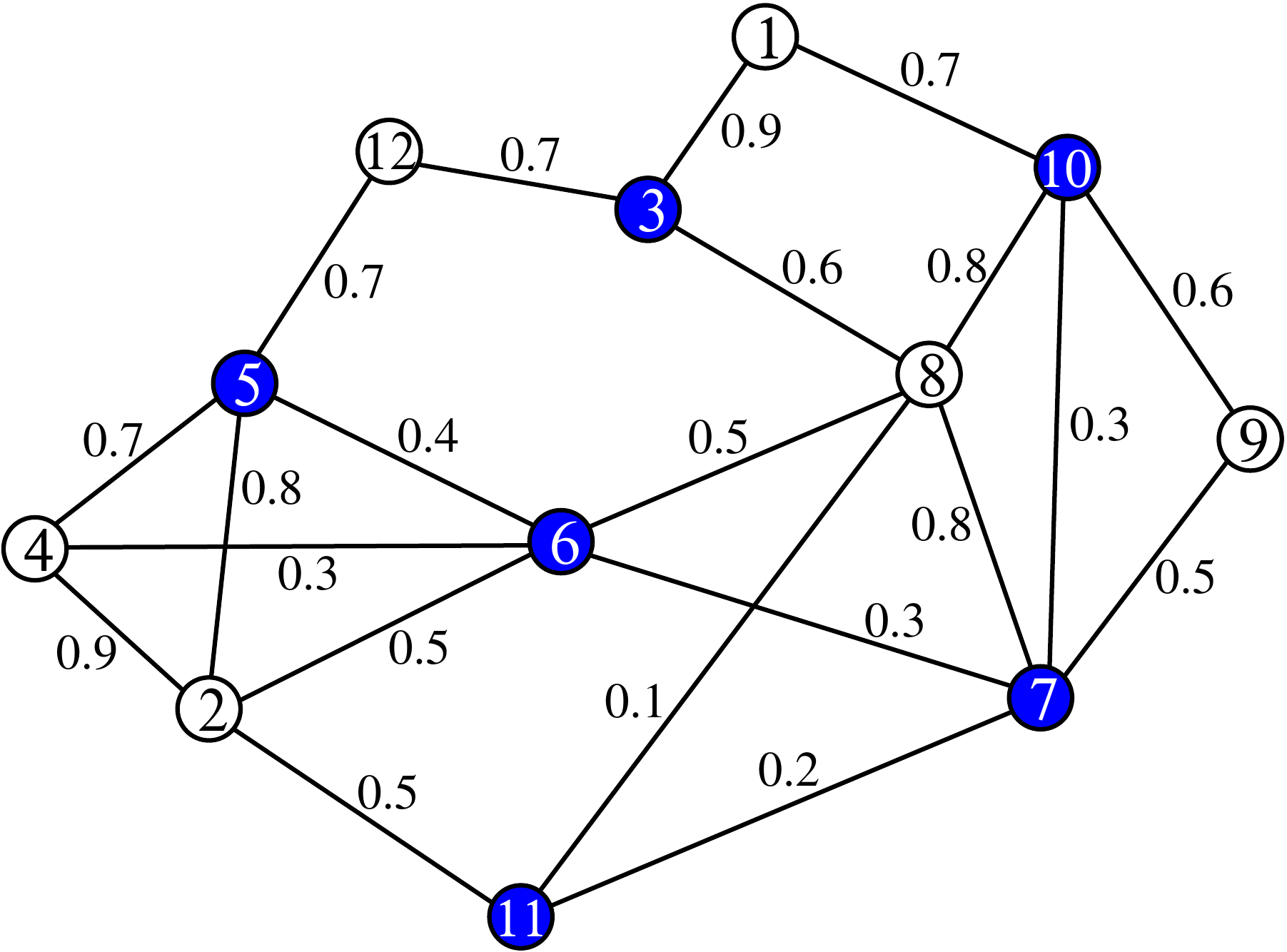}
  \end{center}
  \caption{
    \label{fig:exmpGM}
    An graph with $N=12$ vertices and $M=21$ weighted edges. In this example
    the two weights $w_{i,j}$ and $w_{j,i}$ of each edge $(i,j)$ are equal
    ($w_{i,j}=w_{j,i}$), and the threshold value of each vertex is $\theta=1.0$.
    The vertex set $\Gamma_0  = \{3, 5, 6, 7, 10, 11\}$ is a generalized 
    minimum dominating set for this graph.  The summed
    weight of edges from every
    vertex $j\notin \Gamma_0$ to vertices in $\Gamma_0$
    is equal to or greater than $\theta$.
  }
\end{figure}

We consider a generic graph $G$ formed by $N$ vertices with indices
$i, j, k, \ldots \in \{1, 2, \ldots, N\}$ and $M = (c/2) N$ edges between 
pairs of these vertices (Fig.~\ref{fig:exmpGM}). The constant $c$ is the mean
vertex degree of the graph (on average a vertex is attached with $c$ edges). 
Each edge $(i, j)$ is associated with a pair of non-negative weights
$w_{i,j}$ and $w_{j, i}$ which may or may not be equal. The meaning of the
edge weights depend on the actual context. For example, $w_{i,j}$
may be interpreted as the extent that vertex $i$ represents vertex $j$;
in the symmetric case of $w_{i,j}=w_{j,i}$, we may also interpret $w_{i,j}$
as the similarity between $i$ and $j$. Two vertices $i$ and $j$
are referred to as mutual neighbors if they are connected by an edge $(i,j)$.
The set of neighbors of vertex $i$ is denoted as $\partial i$, i.e.,
$\partial i \equiv \{j\,|\, (i,j)\in G\}$.

Given a graph $G$, we want to construct a vertex set $\Gamma_0$ that is
as small as possible and at the same time is a good representation of all
the other vertices not in this set. 
Let us assign a state
$c_i \in \{0, 1\}$ to each vertex $i$,  $c_i = 1$ if $i \in \Gamma_0$ 
(referred to as being occupied) and $c_i=0$ if $i\notin \Gamma_0$
(referred to as being empty). For each vertex 
$j \notin \Gamma_0$ we require that
$\sum_{i\in \partial j} c_i w_{i,j} \geq \theta$, where $\theta$ is a fixed
threshold value. A vertex $j$ is regarded as being satisfied if it
is occupied ($c_j=1$) or the condition 
$\sum_{i \in \partial j} c_i w_{i,j}\geq \theta$ holds, otherwise it is
regarded as being unsatisfied. Therefore there are $N$ vertex constraints
in the system. A configuration $(c_1, c_2, \ldots, c_N)$
for the whole graph is referred to as a satisfying configuration if and only
if it makes all the vertices to be satisfied (Fig.~\ref{fig:exmpGM}).
Constructing such a generalized MDS $\Gamma_0$, i.e., a satisfying
configuration with the smallest number of occupied vertices, is a
$0$--$1$ integer programming problem, but as it belongs to the 
nondeterministic polynomial-hard (NP-hard) computational complexity class,
no algorithm is guaranteed to solve it in polynomial time. We now seek to
solve it approximately through a statistical physics approach.

Let us introduce a weighted sum $Z(\beta)$ of all the $2^N$ possible 
microscopic configurations $(c_1, c_2, \ldots, c_N)$ as
\begin{equation}
  \label{eq.partition}
  Z(\beta)=\sum_{c_1, \ldots, c_N}  \prod\limits_{j=1}^{N} \biggl[
    \delta_{c_j}^1 e^{-\beta}
    + \delta_{c_j}^0 \Theta\Bigl(\sum\limits_{i\in \partial j} c_i w_{i,j}
    - \theta \Bigr) \biggr]
  \; ,
\end{equation}
where $\delta_{a}^{b}$ is the Kronecker symbol ($\delta_{a}^{b}=1$ if
$a=b$ and $\delta_{a}^{b}=0$ if $a\neq b$), and $\Theta(x)$ is the Heaviside
step function such that $\Theta(x)=0$ for $x<0$ and $\Theta(x)=1$ for $x\geq 0$.
In the statistical physics community, 
$Z(\beta)$ is known as the partition function 
and the non-negative parameter $\beta$ is the inverse temperature.
Notice a configuration $(c_1, c_2, \ldots, c_N)$ has no contribution
to $Z(\beta)$ if it is not a satisfying configuration.
If a configuration satisfies all the vertex constraints, it contributes a
term $e^{-\beta N_1}$ to $Z(\beta)$, where $N_1 \equiv \sum_{i=1}^{N} c_i$ is
the total number of occupied vertices. As $\beta$ increases, satisfying 
configurations with smaller $N_1$ values become
more important for $Z(\beta)$, and at $\beta\rightarrow \infty$ the partition
function is contributed exclusively by the satisfying configurations with 
the smallest $N_1$. For the purpose of constructing a minimum or
close-to-minimum dominating set, we are therefore interested in the 
large-$\beta$ limit of $Z(\beta)$.

\section{Replica-symmetric mean field theory}
\label{sec.mft}

It is very difficult to compute the partition function $Z(\beta)$ exactly,
here we compute it approximately using the replica-symmetric mean 
field theory of statistical physics. This RS mean field theory can be
understood from the angle of Bethe-Peierls approximation 
\cite{Mezard-Parisi-2001,Mezard-Montanari-2009},  it can also be
derived through loop expansion of the partition function 
\cite{Zhou-Wang-2012,Zhou-2015}. 

\subsection{Thermodynamic quantities}

We denote by $q_{j}^{c_j}$ the marginal probability that vertex $j$
is in state $c_j \in \{0, 1\}$. Due to the constraints
associated with vertex $j$ and all its neighboring vertices, the state
$c_j$ is strongly correlated with those of the neighbors.
To write down an approximate expression for $q_{j}^{c_j}$, let us
assume that the states of all the vertices in set $\partial j$ are independent
before the constraint of vertex $j$ is enforced. Under this
Bethe-Peierls approximation we then obtain that
\begin{equation}
  \label{eq.margprob}
  q^{c_j}_j \approx 
  \frac{
    \delta_{c_j}^1 
    e^{-\beta}
    \sum\limits_{\{c_i\,: \,i\in \partial j\}}
    \prod\limits_{i \in \partial j} 
    q^{(c_i, 1)}_{i \rightarrow j}
    +
    \delta_{c_j}^0
    \sum\limits_{\{c_i\,: \,i\in \partial j\}}
    \Theta\Bigl(
    \sum\limits_{i\in \partial j} c_i w_{i,j} - \theta\Bigr)
    \prod\limits_{i \in \partial j}
    q^{(c_i, 0)}_{i\rightarrow j}
  }{
    e^{-\beta}
    \sum\limits_{\{c_i\,: \,i\in \partial j\}}
    \prod\limits_{i \in \partial j} 
    q^{(c_i, 1)}_{i \rightarrow j}
    +
    \sum\limits_{\{c_i\,: \,i\in \partial j\}}
    \Theta\Bigl(
    \sum\limits_{i\in \partial j} c_i w_{i,j} - \theta\Bigr)
    \prod\limits_{i \in \partial j}
    q^{(c_i, 0)}_{i\rightarrow j}
  }
  \; .
\end{equation}
In the above equation,  $q_{i\rightarrow j}^{(c_i, c_j)}$ is the joint probability
that vertex $i$ has state $c_i$ and its neighboring vertex $j$ has state $c_j$
\emph{when the constraint associated with vertex $j$ is not enforced}.
The product $\prod_{i\in \partial j} q_{i\rightarrow j}^{(c_i, c_j)}$ is
a direct consequence of neglecting the correlations among vertices
in $\partial j$ in the absence of vertex $j$'s constraint.
The mean fraction $\rho \equiv N_1 / N$ of occupied vertices is then obtained
through
\begin{equation}
  \label{eq.ocupied}
  \rho =  \frac{1}{N} \sum\limits_{j=1}^{N} q_j^1 \; ,
\end{equation}
This fraction should be a decreasing function of $\beta$.

We can define the free energy of the system as 
$F(\beta)= - \frac{1}{\beta}\ln Z(\beta)$. Within the RS mean field theory
this free energy can be computed through
\begin{equation}
  \label{eq.T_frengy}
  F \equiv N f =\sum\limits_{j=1}^{N} f_j -\sum\limits_{(i,j) \in G}f_{(i,j)} \; ,
\end{equation}
where $f$ is the free energy density; 
and $f_j$ and $f_{(i, j)}$ are, respectively, the free energy contribution
of a vertex $j$ and an edge $(i,j)$:
\begin{subequations}
  \begin{align}
    f_j & =
    -\frac{1}{\beta} \ln \biggl[
      e^{-\beta} \sum\limits_{\{c_i\,: \, i\in \partial j\}}
      \prod\limits_{i \in \partial j}
      q^{(c_i,1)}_{i \rightarrow j} +
      \sum\limits_{\{c_i\,: \, i\in \partial j\}}
      \Theta\Bigl( 
      \sum_{i\in \partial j} c_i w_{i,j} - \theta\Bigr)
      \prod\limits_{i\in \partial j}
      q^{(c_i, 0)}_{i\rightarrow j}
      \biggr] \; ,
    \label{eq.fre_den}
    \\
    f_{(i,j)} & =  
    -\frac{1}{\beta} \ln \biggl[
      \sum\limits_{c_i, c_j} q^{(c_i, c_j)}_{i\rightarrow j}
      q^{(c_j, c_i)}_{j\rightarrow i}
      \biggr] \; .
    \label{eq.fre_den_edge}
  \end{align}
\end{subequations}

The partition function is predominantly contributed by satisfying
configurations with number of occupied vertices $N_1 \approx N \rho$, namely
$Z(\beta)  \approx e^{-\rho \beta N} \Omega(\rho)$ with $\Omega(\rho)$
being the total number of satisfying configurations at occupation density
$\rho$. Then the entropy density $s(\rho) \equiv \frac{1}{N} \ln \Omega(\rho)$
of the system is computed through
\begin{equation}
  s=  ( \rho - f ) \beta \; .
\end{equation}
The entropy density is required to be non-negative by definition. 
If $s(\rho)<0$ as $\rho$ decreases below certain value $\rho_0$, then 
$\Omega(\rho)=e^{N s(\rho)} \rightarrow 0$ suggests that there is no
satisfying configurations with $\rho< \rho_0$. We therefore take the
value $\rho_0$ as the fraction of vertices contained in a minimum
dominating set.

\subsection{Belief-propagation equation}

We need to determine the probabilities $q_{i\rightarrow j}^{(c_i, c_j)}$ to
compute the thermodynamic densities $\rho$, $f$, and $s$. Following the
Bethe-Peierls approximation and similar to Eq.~(\ref{eq.margprob}),
$q_{i\rightarrow j}^{(c_i, c_j)}$ is self-consistently determined through
\begin{subequations}
  \label{eq.BP}
  \begin{align}
    q_{i\rightarrow j}^{(0,0)} &=
    \frac{1}{z_{i \rightarrow j}}
    \sum\limits_{\{c_k\,:\,k\in \partial i\backslash j\}}
    \Theta\Bigl(\sum\limits_{k\in \partial i\backslash j}
    c_k w_{k,i} - \theta\Bigr) \prod\limits_{k\in \partial i\backslash j}
    q_{k\rightarrow i}^{(c_k, 0)} \; , 
    \\
    q_{i\rightarrow j}^{(0,1)} & =\frac{1}{z_{i \rightarrow j}}
    \sum\limits_{\{c_k\,:\,k\in \partial i\backslash j\}}
    \Theta\Bigl(w_{j,i}+\sum\limits_{k\in \partial i\backslash j}
    c_k w_{k,i} - \theta\Bigr) \prod\limits_{k\in \partial i\backslash j}
    q_{k\rightarrow i}^{(c_k, 0)} \; ,
    \\
    q_{i\rightarrow j}^{(1,0)} = q_{i\rightarrow j}^{(1,1)} & =
    \frac{1}{z_{i\rightarrow j}}
    e^{-\beta} 
    \prod_{k\in \partial i \backslash j} \Bigl[ q_{k \rightarrow i}^{(1,1)}
      + q_{k \rightarrow i}^{(0,1)} \Bigr]\; ,
  \end{align}
\end{subequations}
where $\partial i \backslash  j$ is the subset of $\partial i$ with vertex
$j$ being deleted, and $z_{i\rightarrow j}$ is a normalization constant.
Equation (\ref{eq.BP}) is called a belief-propagation (BP) equation in the
literature.
To find a solution to Eq.~(\ref{eq.BP}) we iterate this equation on
all the edges of the input graph $G$ (see, for example,
\cite{Zhao-Habibulla-Zhou-2015,Habibulla-Zhao-Zhou-2015} or \cite{Zhou-2015}
for implementing details). However convergence is not guaranteed to
achieve.   If the reweighting parameter $\beta$ is small
this BP iteration quickly reaches a fixed point; 
while at large values of $\beta$ we notice that 
it usually fails to converge (see next subsection). 

\subsection{Results on Erd\"os-R\'enyi random graphs}

We first apply the RS mean field theory to Erd\"os-R\'enyi (ER) random graphs. 
To generate an ER random graph, we select $M$ different pairs of edges 
uniformly at random from the whole set of $N (N-1)/2$ vertex pairs and then
connect each selected pair of vertices by an edge. For $N$ sufficiently large
there is no structural correlations in such a random graph, and the typical
length of a loop in the graph diverges with $N$ in a logarithmic way. 

\begin{figure}
  \centering
  \includegraphics[angle=270,width=1.0 \textwidth]{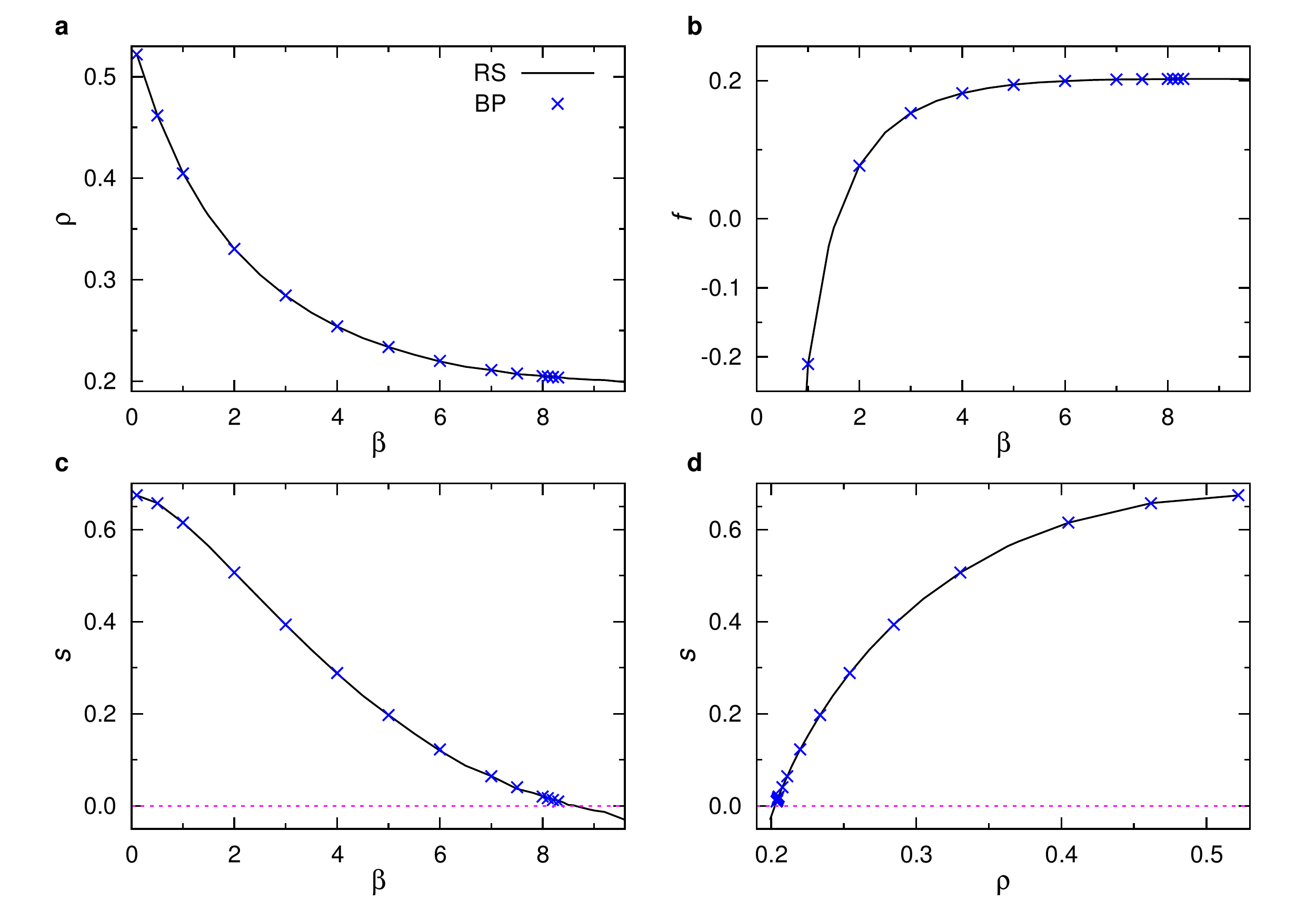}
  \caption{
    Replica-symmetric mean field results on ER random networks of mean vertex
    degree $c=10.0$. The symmetric edge weights are drawn from the set
    $\{0.4, 0.5, 0.6, 0.7, 0.8, 0.9, 1.0\}$ and the vertex threshold value is
    $\theta=1$. The cross symbols (BP) are results obtained by
    belief-propagation on a single graph instance of size $N=10^5$,
    while the solid lines (RS) are ensemble-averaged results obtained by
    population dynamics simulations. 
    (a) Occupation density $\rho$ versus inverse temperature $\beta$;
    (b) free energy density $f$ versus $\beta$; (c) entropy density $s$
    versus $\beta$; (d) entropy density $s$ as a function of $\rho$ obtained 
    by combining data of (a) and (c).
  }
  \label{fig.100kER}
\end{figure}

If the two edge weights of every edge $(i,j)$ are equal to the vertex
threshold value $\theta$ ($w_{i, j}=w_{j,i}=\theta$), the generalized
MDS problem reduces to the conventional MDS problem on an undirected graph,
which has been successfully treated in \cite{Zhao-Habibulla-Zhou-2015}. For
example, for ER random graphs with mean vertex degree $c=10.0$ the MDS 
relative size is $\rho_0 \approx 0.120$ \cite{Zhao-Habibulla-Zhou-2015}.
On the other hand, if the two edge weights of every edge are strongly 
non-symmetric such that either $w_{i,j}=\theta$ and $w_{j,i}=0$ (with
probability $1/2$) or $w_{i,j}=0$ and $w_{j,i}=\theta$ (also with probability
$1/2$), the generalized MDS problem reduces to the
conventional MDS problem on a directed graph, which again has been successfully
treated in \cite{Habibulla-Zhao-Zhou-2015} (e.g., at $c=10.0$ the MDS relative
size is $\rho_0 \approx 0.195$). 

In this paper, as a particular example, we consider a distribution of edge
weights with the following properties: (1) the weights of every edge $(i,j)$ 
are symmetric, so $w_{i, j}=w_{j,i}$; (2) the edge weights of different edges 
are not correlated but completely independent; (3) for each edge $(i, j)$ its
weight $w_{i,j}$ is assigned the value $0.4 \theta$ or $1.0 \theta$ with
probability $1/12$ each and assigned values in the set 
$\{0.5 \theta, 0.6 \theta, 0.7 \theta, 0.8 \theta, 0.9\theta\}$
with equal probability $1/6$ each.

The BP results on the occupation density $\rho$, the free energy density $f$,
and the entropy density $s$ are shown in Fig.~\ref{fig.100kER} for a single 
ER random graph of $N=10^5$ vertices and mean degree $c=10.0$.
The BP iteration for this this graph instance is convergent for 
$0 \leq \beta \leq 8.3$.
The occupation density $\rho$ and the entropy density $s$ both decrease with 
inverse temperature $\beta$.
The entropy density as a function of occupation density, $s(\rho)$, approaches
zero at $\rho = \rho_0 \approx 0.202$, indicating there is no satisfying
configurations at occupation density $\rho < \rho_0$. 
The BP results therefore predict that a MDS for this problem instance
must contain at  least $0.202 N$  vertices.

We can also obtain RS mean field results on the thermodynamic densities
by averaging over the whole ensemble of ER random graphs 
(with $N\rightarrow \infty$ and fixed
mean vertex degree $c$). This is achieved by population dynamics
simulations \cite{Mezard-Parisi-2001}. We store a
population of probabilities $\{q_{i\rightarrow j}^{(c_i, c_j)}\}$ and update
this population using Eq.~(\ref{eq.BP}), and at the same time compute
the densities of thermodynamic quantities. A detailed description on 
the implementation can be found in section 4.3 of
\cite{Zhao-Habibulla-Zhou-2015}. 
The ensemble-averaged results for the ER random network ensemble of
$c=10.0$ and $N\rightarrow \infty$ are also shown in Fig.~\ref{fig.100kER}.
These results are in good agreement with the BP results obtained on the
single graph instance. 

\begin{figure}
  \begin{center}
    \includegraphics[angle=270, width=0.6 \textwidth]{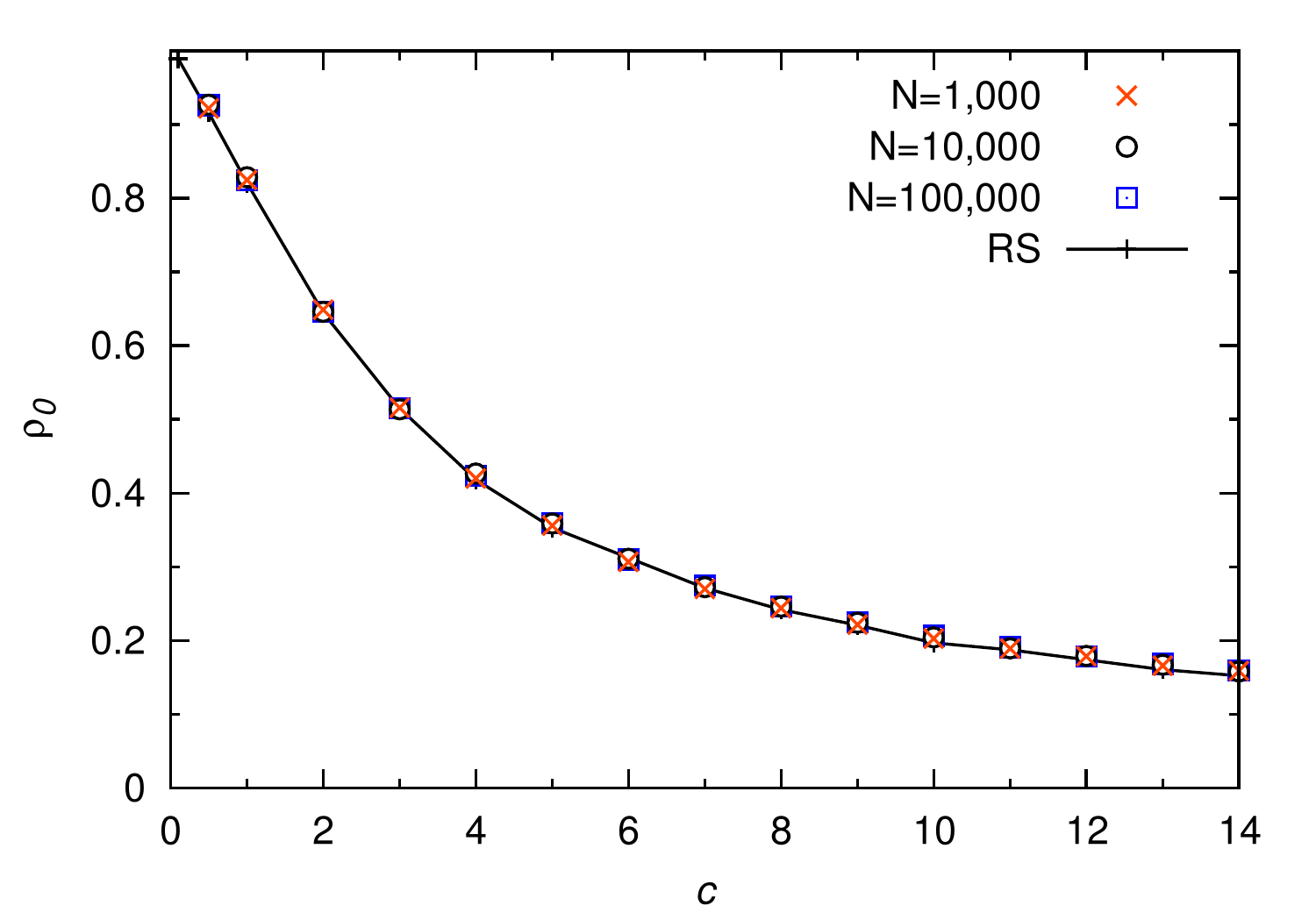}
  \end{center}
  \caption{
    \label{fig.bpd_n}
    The relative size $\rho_0$ of minimum dominating sets for
    ER random graphs of mean vertex degree $c$. The edge weight distribution
    for these random graphs are the same as that of Fig.~\ref{fig.100kER},
    and the vertex threshold value $\theta=1.0$. 
    The solid-line connected plus symbols are the predictions of the
    RS mean field theory, while the results obtained by the BPD algorithm
    at $\beta=8.0$  are drawn as cross symbols (for graph size $N=10^3$), 
    circles ($N=10^4$), and squares ($N=10^5$). 
    Each BPD data point is the result of a single run on one graph instance.
  }
\end{figure}

Through the RS population dynamics simulations we can estimate the
ensemble-averaged value of $\rho_0$ (the minimum fraction of occupied
vertices) by the equation $s(\rho_0) = 0$. The value of $\rho_0$ obtained
in such a way decreases with mean vertex degree $c$ continuously, 
see Fig.~\ref{fig.bpd_n} (solid line).

\section{Belief-propagation-guided decimation algorithm}
\label{sec.BPD}

For $\beta$ sufficiently large, the marginal occupation probability $q_j^{c_j}$
obtained by Eq.~(\ref{eq.margprob}) tells us the likelihood of each vertex $j$
to belong to a minimum dominating set. This information can serve as a guide
for constructing close-to-minimum dominating sets.  Based on the BP equation
(\ref{eq.margprob}) we implement a simple belief-propagation-guided 
decimation (BPD) algorithm as follows. Starting from an input graph $G$ and an
empty vertex set $\Gamma$, at each step we (1) iterate the BP equation for
a number of repeats and then estimate the occupation probability
$q_j^{1}$ for all the vertices $j$ not in $\Gamma$; and (2) add a tiny fraction 
(e.g., $1\%$) of those vertices $j$ with the highest values of $q_j^{1}$ 
into the set $\Gamma$ and set their state to be $c_j=1$; (3) then simplify 
the graph and repeat the operations (1)--(3) on the simplified graph, 
until $\Gamma$ becomes a dominating set.

The detailed implementation of this BPD algorithm is the same as described in
section 5 of \cite{Zhao-Habibulla-Zhou-2015}. Here we only need to emphasize
one new feature: after a vertex $i$ is newly occupied, the
threshold  value  (say $\theta_j$) of every neighboring vertex $j$ should be
updated as $\theta_j \leftarrow (\theta_j - w_{i,j})$, and if
this updated $\theta_j$ is non-positive then vertex $j$ should be regarded
as being satisfied.

For the same graph of Fig.~\ref{fig.100kER}, a single trial of this BPD
algorithm at $\beta=8.0$ results in a dominating set of size $21009$,
which is very close to the predicted MDS size by the RS mean field theory.
Equally good performance of the BPD algorithm is also achieved on other
ER random graphs with mean vertex degree $c$ ranging from $c=0.5$ to $c=14$
(see Fig.~\ref{fig.bpd_n}), suggesting that the BPD algorithm is able to 
construct a dominating set which is very close to a MDS.
We emphasize that in the BPD algorithm we do not require
  the BP iteration to converge.

\section{Application: Automatic text summarization}
\label{sec.app}

Automatic text summarization is an important issue in the research field
of natural language processing \cite{Mani-1999}. One is faced with the
difficult task of constructing a set of sentences to
summarize a text document (or a collection of text documents) in a most
informative and efficient way. Here we extend the initial idea of
Shen and Li \cite{Shen-Li-2010} and consider this information retrieval
problem as a generalized minimum dominating set problem.

We represent each sentence of an input text document as a vertex and 
connect two vertices (say $i$ and $j$) by an weighted edge, with
the symmetric edge weight $w_{i,j}$ ($=w_{j,i}$) being equal to
the similarity of the two corresponding sentences. Before computing
the edge weight a pre-treatment is applied to all the sentences to
remove stop-words (such as `a', `an', `at', `do', `but', `of', `with')
and to transform words to their prototypes according to the
WordNet dictionary \cite{Fellbaum-1998} (e.g., `airier' $\rightarrow$
`airy', `fleshier' $\rightarrow$ fleshy, `are' $\rightarrow$ `be', 
`children' $\rightarrow$ child, `looking' $\rightarrow$ `look').
There are different ways to measure sentence similarity, here we consider
a simple one, the cosine similarity \cite{Singhal-2001}. To compute the
cosine similarity, we map each sentence $i$ to a high-dimensional vector
$\vec{S}_i$, the $k$-th element of which is just the number of times the
$k$-th word of the text appears in this sentence. 
Then the edge weight between
vertices $i$ and $j$ is defined as
\begin{equation}
  w_{i,j} = \frac{ \vec{S}_i \cdot \vec{S}_j}{
    \sqrt{ \vec{S}_i \cdot \vec{S}_i} \sqrt{\vec{S}_j \cdot \vec{S}_j}}
  \; . 
\end{equation}
To give a simple example, let us consider a document with only two 
sentences `Tom is looking at his children with a smile.' and
`These children are good at singing.'. The  word set of this document
is \{Tom, be, look, child, smile, good, sing\}, and
the vectors for the two sentences are
$\vec{S}_1=(1, 1, 1, 1, 1, 0, 0)$ and
$\vec{S}_2= (0, 1, 0, 1, 0, 1, 1)$, respectively. 
The cosine similarity $w_{1 2}$
between these two sentences is then $w_{1 2}=\frac{2}{\sqrt{5} \sqrt{4}}
\approx 0.447$.

We first test the performance of the BPD algorithm on $32$ short English text 
documents of different lengths (on average a document has $17.7$ sentences).
We compare the outputs from the BPD algorithm with the key sentences 
manually selected by the first author.
For each text document we denote by
$B$ and $\tilde{B}$ the set of key sentences selected by human inspection 
and by the algorithm, respectively. 
On average the set $B$ of human inspection 
contains a fraction $\rho=0.226$ of the sentences
in the input text document.
Then we define
the coverage ratio $R_{cov}$ and the difference ratio $R_{dif}$ between 
$B$ and $\tilde{B}$ as
\begin{equation}
  \label{eq.RRmeasure}
  R_{cov} =\frac{\bigl| B \cap \tilde{B}\bigr|}{\bigl| B \bigr|}\;, 
  \quad \quad\quad
  R_{dif}=\frac{\bigl| \tilde{B}- B \bigl|}{\bigl| \tilde{B} \bigr|} \; ,
\end{equation}
where $(\tilde{B}-B)$ denotes the set of sentences belonging to $\tilde{B}$ but
not to $B$. The ratio $R_{cov}$ quantifies the probability of a manually 
selected key sentence also being selected by the algorithm, while the ratio
$R_{dif}$ quantifies the extent that a sentence selected by the algorithm does
not belong to the set of manually selected key sentences.

\begin{table}
  \centering
  \caption{Averaged performances of the BPD algorithm ($\beta=8.0$),
    the PR (PageRank) algorithm, and the AP (affinity-propagation) algorithm
    on $32$ English text documents (average number of sentences per document
    $17.7$). For BPD the vertex threshold is set to
    $\theta=0.6$ (BPD$_{0.6}$), $\theta=0.8$ (BPD$_{0.8}$) and
    $\theta=1.0$ (BPD$_{1.0}$). For PR the fraction of sentences selected
    is $25\%$ (PR$_{25\%}$), $30\%$ (PR$_{30\%}$), and $40\%$ (PR$_{40\%}$).
    For AP the adjustable parameter is set to be $w_{i,i}=0.0$ (AP$_{0.0}$) and
    $w_{i,i}=0.2$ (AP$_{0.2}$). $\rho$ is the fraction of representative
    sentences chosen by the algorithm, and $R_{cov}$ and $R_{dif}$ are
    two performance measures defined by Eq.~(\ref{eq.RRmeasure}).
    The average fraction of representative sentences constructed by human 
    inspection is $\rho=0.226$.
    \label{tab:result1}
  }
\vskip 0.2cm
  \begin{tabular}{l|ccc|ccc|cc}
    \hline
    & BPD$_{0.6}$ & BPD$_{0.8}$ & BPD$_{1.0}$
    & PR$_{25\%}$ & PR$_{30\%}$ & PR$_{40\%}$ & AP$_{0.0}$ & AP$_{0.2}$  \\
    \hline
    $\rho$ & $0.44$ & $0.48$ & $0.56$ & $0.27$ & $0.32$ & $0.42$ & 
    $0.17$ & $0.39$ \\
    $R_{cov}$ & $39.9\%$ & $47.2\%$ & $49.6\%$ & $30.0\%$ & $41.7\%$
    & $50.6\%$ & $15.6\%$ & $39.3\%$ \\
    $R_{dif}$ & $79.4\%$ & $78.6\%$ & $80.2\%$ & $74.2\%$ & $71.4\%$ 
    & $72.2\%$ & $76.3\%$ & $77.4\%$ \\
    \hline
  \end{tabular}
\end{table}

We also apply two other summarization algorithms to the same set of
text documents, one is the PageRank (PR) algorithm 
\cite{Brin-Page-1998,Mihalcea-Tarau-2004,Erkan-Radev-2004}, and 
the other is the
Affinity-Propagation (AP) algorithm
\cite{Frey-Dueck-2007}. 
PageRank is based on the idea of random walk on a graph, and
it offers an efficient way of measuring vertex significance.
The importance $P_i$ of a vertex $i$ is determined by the following 
self-consistent equation
\begin{equation}
  P_i=(1-p) \frac{1}{N} + p * \sum_{j\in\partial i} P_j \frac{w_{j,i}}{
  \sum_{k \in \partial j} w_{j,k}} \; ,
\end{equation}
where $p$ is the probability to jump from one vertex to a neighboring vertex
(we set $p=0.85$ following \cite{Brin-Page-1998}).
Those vertices $i$ with high values of $P_i$ are then selected as the
representative vertices.

On the other hand, Affinity-Propagation is a clustering 
algorithm: each vertex either selects a neighboring vertex as its
exemplar or serves as an exemplar for some or all of its neighbors
\cite{Frey-Dueck-2007}.
For any pair of vertices $i$ and $j$, the responsibility $r_{i,j}$ 
of $j$ to $i$ and the availability $a_{i,j}$ of $j$ to $i$ are determined by
the following set of iterative equations: 
\begin{subequations}
  \label{eq.AP}
  \begin{align}
    r_{i,j} & = w_{i,j} - \max_{k \neq j} \bigl\{a_{i,k}+w_{i,k}\bigr\} \; ,
    \label{eq.APa}\\
    a_{i,j} & = \min\Bigl[ 0, \; 
      r_{j,j} + \sum_{k \neq  i,j} \max\bigl[0,\, r_{k,j}\bigr] \Big] \; , \\
    a_{j,j} &= \sum_{i \neq j} \max \bigl[ 0,\; r_{i, j}\bigr] \; .
  \end{align}
\end{subequations}
In Eq.~(\ref{eq.APa}) $w_{i,j}$ is the weight of edge $(i,j)$ for $i\neq j$,
and $w_{i,i}$ is an adjustable parameter which affects the final number of
examplars. We iterate the AP equation (\ref{eq.AP}) on the sentence graph 
starting from the initial condition of $r_{i,j}=a_{i,j}=0$ and, after
convergence is reached,
then consider all the vertices $i$ with positive values of $(r_{i,i}+a_{i,i})$
as the examplar vertices.

For the $32$ short text documents used in our preliminary test, the comparative
results of Table~\ref{tab:result1} do not distinguish much the three heuristic
algorithms, yet it appears that PageRank performs slightly
  better than BPD and AP. When the fraction of extracted sentences is
 $\rho=0.42$, the
coverage ratio reached by PR is $R_{cov}=0.51$ and the
difference ratio is $R_{dif}=0.72$, while $R_{cov}=0.40$ and
$R_{dif}=0.79$ for BPD at $\rho=0.44$ and $R_{cov}=0.39$ and
$R_{dif}=0.77$ for AP at $\rho=0.39$.
 
We then continue to evaluate the performance of the belief-propagation 
approach on a benchmark set of longer text documents, namely the DUC 
(Document Understanding Conference) data set used in \cite{Erkan-Radev-2004}.
We examine a total number of $533$ text documents from the DUC 2002 directory
\cite{DUC}. The average number of sentences per
document is about $28$ and the average number of words per sentence is about
$20$.

The DUC data set offers, for each of these text documents, 
two sets $B$ of representative sentences chosen by two human experts,
the total number of words in such a set $B$ being $\approx 100$.
The PageRank
algorithm (PR$^{100}$) and one version of the BPD algorithm
(BPD$_{\theta}^{100}$, $\theta=0.6$ or $\theta=1.0$) also construct a
set $\tilde{B}$ of sentences for each
of these documents under the constraint that the total
number of words in $\tilde{B}$ should be about $100$. In another version of
the BPD algorithm (BPD$_{\theta}$) the restriction on the words number in
$\tilde{B}$ is removed. We follow the DUC convention and 
use the toolkit ROUGE \cite{ROUGE} to evaluate the agreement between
$\tilde{B}$ and $B$ in terms of Recall, Precision, and F-score:
\begin{subequations}
  \begin{align}
   \mathrm{Recall} 
    & = 
    \frac{\sum\limits_{\mathrm{word}\in B} 
      \min\bigl[C(\mathrm{word}), \tilde{C}(\mathrm{word}) \bigr]}
         {\mathrm{WordsNum}(B)} \; ,
         \\
    \mathrm{Precision} & = 
         \frac{\sum\limits_{\mathrm{word}\in B} 
           \min\bigl[C(\mathrm{word}), \tilde{C}(\mathrm{word}) \bigr]}
              {\mathrm{WordsNum}(\tilde{B})} \; ,
              \\
         \mathrm{Fscore} &=
              \frac{2 \times \mathrm{Precision} \times \mathrm{Recall}}
                   {\mathrm{Precision}+\mathrm{Recall}} \; .
  \end{align}
\end{subequations}
where $C(\mathrm{word})$ is the total number of times a given word appears
in the summary $B$, and $\tilde{C}(\mathrm{word})$ is the number of times
this word appears in the summary $\tilde{B}$; $\mathrm{WordsNum}(B)$
is the total number of words in the summary $B$ and similarly for
$\mathrm{WordsNum}(\tilde{B})$.

\begin{table}
  \centering
  \caption{
    Averaged performances of the BPD algorithms BPD$_{\theta}^{100}$
    ($\theta=0.6$ or $\theta=1.0$) and
    BPD$_{\theta}$ ($\theta=1.0$) and the PageRank algorithm PR$^{100}$ 
    on the $533$ text documents of DUC 2002 \cite{DUC}. The Precision, Recall,
    and F-score values are obtained by averaging over the results of
    individual text documents. The inverse temperature of BPD is fixed to
    be $\beta=8.0$.
    \label{tab.DUC}
  }
  \vskip 0.2cm
  \begin{tabular}{l|c|cc|c}
    \hline
    & PR$^{100}$ & BPD$_{0.6}^{100}$ & BPD$_{1.0}^{100}$ &  BPD$_{1.0}$ \\
    \hline
    Recall  & $0.455$ & $0.249$ & $0.264$ & $0.727$ \\
    Precision & $0.407$ & $0.396$ & $0.410$ & $0.256$ \\
    Fscore & $0.429$ & $0.303$ & $0.318$ & $0.359$ \\
    \hline
  \end{tabular}
\end{table}

The comparative results for the DUC 2002 data set are shown in 
Table~\ref{tab.DUC}. We notice that BPD$_{1.0}$ ($\theta=1.0$) has the
highest Recall value of $0.727$, namely the summary obtained by this
algorithm contains most of contents in the summary of human experts, but its
Precision value of $0.256$ is much lower than that of the PR$^{100}$ algorithm,
indicating that the BPD algorithm add more sentences into the summary than
the human experts do. In terms of the F-score which balances Recall and
Precision (the last row of Table~\ref{tab.DUC}) we conclude that
PageRank also performs a little bit better than BPD for the DUC 2002 benchmark.

The generalized MDS model for the text summarization problem aims at
a complete coverage of an input text document.
It is therefore natural that the summary constructed by BPD contains
more sentences than the summary constructed by the human experts (which may
only choose the sentences that best summarize the key points of a text
document). All the tested documents in the present work
are rather short, which may make the
advantages of the BPD message-passing algorithm difficult to be
manifested. More work needs to be done to test the performance of
the BPD algorithm on very long text documents.

\section{Outlook}
\label{sec.conc}

In this paper we presented a replica-symmetric mean field theory for the
generalized minimum dominating set problem, and we considered the task of
automatic text summarization as such a MDS problem and applied the BPD
message-passing algorithm to construct a set of representative sentences
for a text document. When tested on a set of short text documents the BPD
algorithm has comparable performance as the PageRank and the
Affinity-Propagation algorithms. 
We feel that the BPD approach will be most powerful for extracting
sentences out of lengthy text documents (e.g., scientific papers
containing thousands of sentences).   We hope that our work will stimulate
further efforts on this important application.

The belief-propagation based method for the automatic text summarization 
problem might be improved in various ways. For example, it may not be 
necessary to perform the decimation step, rather one may run BP on the input
sentence graph until convergence (or for a sufficient number of rounds)
and then return an adjustable fraction
$\rho$ of the sentences $i$ according to their estimated 
occupation probabilities $q_i^1$.

\begin{figure}
  \begin{center}
    \includegraphics[width=0.5\textwidth]{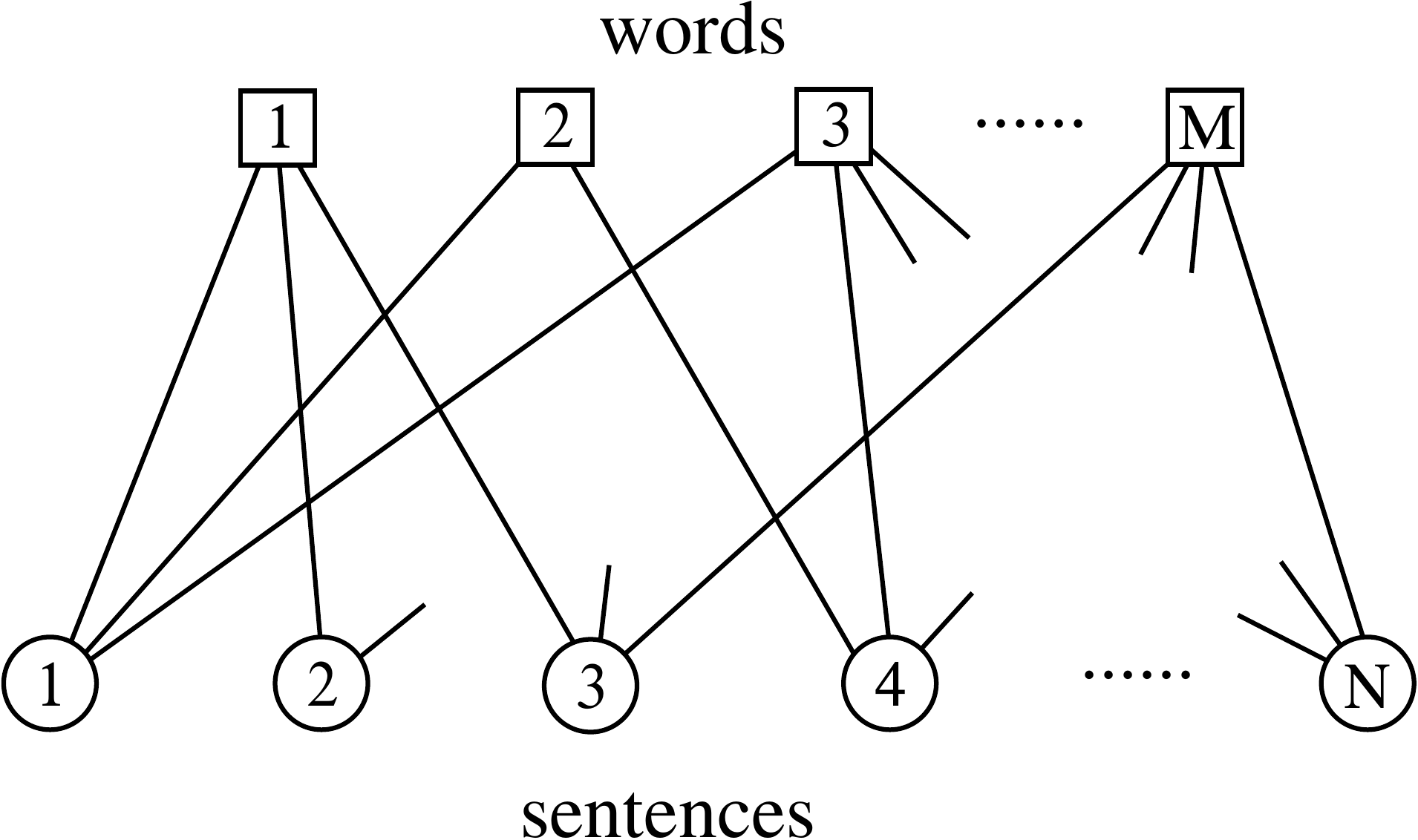}
  \end{center}
  \caption{
    \label{fig:wordsentence}
    The word--sentence graph representation for a text document. The 
    $M$ words and $N$ sentences of an input text document are denoted by 
    squares and circles, respectively, and a link between a word $a$ and a
    sentence $i$
    is drawn if and only if word $a$ appears in sentence $i$.
    To get a set $\Gamma$ of representative sentences we may require that
    each word must be connected to at least $n$ ($n\geq 1$) sentences of
    the set $\Gamma$.
  }
\end{figure}

One may also convert the text summarization problem to other 
generalized MDS problems. A particularly simple
but potentially useful one can be constructed as follows: we first
construct a bi-partite graph formed by words, sentences, and the links
between words and sentences (see Fig.~\ref{fig:wordsentence}); we then
construct a minimum-sized dominating set of sentences $\Gamma$ such that 
every word of the whole bipartite graph must appear in at least
$n$ ($n\geq 1$) of the sentences
of $\Gamma$. Such a generalized MDS problem can be studied 
by slightly modifying the BP equation Eq.~(\ref{eq.BP}). We notice that this
alternative construction has the advantage of encouraging
diversity in the selected representative sentences. 

\ack

We thank Jin-Hua Zhao and Yusupjan Habibulla for helpful discussions. This
research is partially supported by the National Basic Research Program of 
China (grant number 2013CB932804) and by the National Natural Science
Foundation of China (grand numbers 11121403 and 11225526).

\end{document}